# Pinalites: Optical properties and Quantum Magnetism of Heteroanionic $A_3MO_5X_2$ Compounds

Daigorou Hirai*,†

†Department of Applied Physics, Nagoya University, Nagoya 464–8603, Japan

**ABSTRACT:** Heteroanionic compounds, which contain two or more types of anions, have emerged as a promising class of materials with diverse properties and functionalities. In this paper, I review the experimental findings on $Ca_3ReO_5Cl_2$ and related compounds that exhibit remarkable pleochroism and novel quantum magnetism. I discuss how the heteroanionic coordination affects the optical and magnetic properties by modulating the d-orbital states of the transition metal ions and then compare these materials with other heteroanionic and monoanionic compounds and highlight the potential of $A_3MO_5X_2$ materials for future exploration of materials and phenomena.

## 1. INTRODUCTION

Heteroanionic compounds, which contain two or more types of anions, have unique crystal structures and properties that are not found in monoanionic compounds such as oxides[1]. Heteroanionic compounds have been explored less than monoanionic compounds, but recent research has revealed various physical and chemical phenomena, such as superconductivity in LaOFePn (Pn = P, As)[2,3], $Li_xHfNCl$[4] and $(Ca_{1-x}Na_x)CuO_2Cl_2$[5], high dielectric constants in $AMO_2N$ (A = Ba, Sr, Ca; M= Ta, Nb)[6], ion conductivity in LaOCl[7] and $La_2LiHO_3$[8], catalytic activity in $BaTiO_{3-x}H_x$[9] and TaON[10], and the thermoelectric properties of MCuOCh (M = Bi, La; Ch = S, Se, Te) [11,12].

One of the heteroanionic compounds that has attracted much attention is the oxychloride $Ca_3ReO_5Cl_2$ (known as CROC), which exhibits a remarkable optical property called pleochroism; that is, the color changes depending on the viewing direction and the polarization of the incident light. While pleochroism is an optical property found in a wide variety of minerals, CROC shows a dramatic change in color, from green to yellow to red, depending on the direction of polarization. CROC also shows novel quantum magnetism as a result of magnetic frustration. These properties are related to the heteroanionic coordination of the transition metal ion, which is influenced by the presence of both oxygen and chlorine anions. Several other compounds with similar characteristics have been synthesized and investigated, and they have the general formula $A_3MO_5X_2$.

In this review, I summarize the crystal structural features and the recent studies on the optical and magnetic properties of $A_3MO_5X_2$ compounds, with a focus on the role of the heteroanionic coordination. I also compare these compounds with other heteroanionic and monoanionic compounds and discuss the future prospects of Pinalite family ($A_3MO_5X_2$ materials). In the field of optical properties and quantum magnetism, where systematic material exploitation of heteroanionic compounds has been less developed, Pinalite compounds are a good model case for the usefulness of heteroanionic nature.

### 1.1 Pinalits: $A_3MO_5X_2$ COMPOUDS

$A_3MO_5X_2$ compounds are a series of heteroanionic compounds with different combinations of A, M, and X ions. One of this group of compounds, $Pb_3WO_5Cl_2$, was first found in the mineral pinalite in Arizona[14,15], and, here, the $A_3MO_5X_2$ compounds are named "Pinalites". Table 1 lists the reported compounds so far. In these materials, alkaline earth metals and Pb occupy the A site as $A^{2+}$ ions, early transition metals (Mo, Re, and W) occupy the M site as $M^{6+}$ ions, and halogens Cl and Br occupy the X site as $X^-$ ions.

Since the discovery of CROC, compounds with $Re^{6+}$ ions in the $5d^1$ electron configuration at the M site have been studied extensively, and their optical and magnetic properties have been of great interest. $A_3MO_5X_2$ compounds have two different types of crystal structures, as shown in Fig. 1a: the $Ca_3WO_5Cl_2$-type structure in the *Pnma* space group having A = Ca, and a $Pb_3WO_5Cl_2$-type structure in the *Cmcm* space group (or *Amam*, which has the same symmetry but different axis settings) for other compounds. The difference between the two structures is the distortion caused by the smaller ionic size of $Ca^{2+}$ compared to other $A^{2+}$ ions[13].

### 1.2 $Pb_3WO_5Cl_2$-TYPE STRUCTURE

Most compounds in the $A_3MO_5X_2$ family crystallize in the $Pb_3WO_5Cl_2$-type structure. The $Pb_3WO_5Cl_2$-type structure can be viewed as a layered structure with alternating layers of $A_3MO_5$ and $X_2$. The $A_3MO_5$ layer is derived from the $[Pb_2O_2]$ layer in the structure of the mineral litharge, α-PbO[16], which is a common motif in lead halides[17]. The $[Pb_2O_2]$ layer consists of a square lattice of oxide atoms and a checkerboard arrangement of Pb atoms above and below the center of the oxygen squares. The $A_3MO_5$ layer also has a square lattice of oxide atoms at the O1 sites, but this lattice is corrugated. As shown in Fig. 1b, three-fourths of the Pb atoms in the $[Pb_2O_2]$ layer are replaced by A atoms in the $A_3MO_5$ layer, and the remaining one-fourth are replaced by

M-O units. The corrugations of the oxygen square lattice in the $A_3MO_5$ layer are due to the different sizes and shapes of the M-O units and the A cations.

The $X_2$ layers between the $A_3MO_5$ layers also form a square lattice that matches the opposite side of the protruding A ion and the M-O unit. There are two X sites: X1 facing the A ion and X2 facing the M-O unit. The distance between X2 and the M-O unit is slightly shorter than that between X1 and the A ion, resulting in a slight distortion of the X square lattice.

**Table 1.** Compounds with the general chemical formula $A_3MO_5X_2$. The space group for $Pb_3WO_5Cl_2$-type structure is standardized to *Cmcm*. ATL stands for anisotropic triangular lattice.

| Chemical formula | Space group | Electronic configuration | Lattice constants (Å) | Unit cell volume (Å$^3$) | Comments | Ref. |
|---|---|---|---|---|---|---|
| $Ca_3ReO_5Cl_2$ | *Pnma* (#62) | 5d$^1$ | $a$ = 11.8997(2)<br>$b$ = 5.5661(1)<br>$c$ = 11.1212(2) | $V$ = 736.61(2) | Pleochroism<br>ATL magnet ($J'/J$ = 0.25)<br>$T_N$ = 1.13 K | 18,19 |
| $Ca_3ReO_5Br_2$ | *Pnma* (#62) | 5d$^1$ | $a$ = 12.1219(14)<br>$b$ = 5.6447(7)<br>$c$ = 11.2205(10) | $V$ = 767.76(15) | ATL magnet ($J'/J$ = 0.25)<br>$T_N$ = 1.15 K | 20 |
| $Ca_3WO_5Cl_2$ | *Pnma* (#62) | 5d$^0$ | $a$ = 11.820(2)<br>$b$ = 5.587(1)<br>$c$ = 11.132(1) | $V$ = 735.2 | Charge transfer luminescence | 21,22 |
| $Sr_3ReO_5Cl_2$ | *Cmcm* (#63) | 5d$^1$ | $a$ = 5.6492(3)<br>$b$ = 13.1886(6)<br>$c$ = 11.1144(5) | $V$ = 828.09(7) | ATL magnet ($J'/J$ = 0.35)<br>$T_N$ < 1.8 K | 23 |
| $Sr_3ReO_5Br_2$ | *Cmcm* (#63) | 5d$^1$ | $a$ = 5.7145(3)<br>$b$ = 13.3445(7)<br>$c$ = 11.3044(6) | $V$ = 862.05(8) | ATL magnet ($J'/J$ = 0.25)<br>$T_N$ < 1.8 K | 20 |
| $Ba_3ReO_5Cl_2$ | *Cmcm* (#63) | 5d$^1$ | $a$ = 5.79424(18)<br>$b$ = 13.9508(4)<br>$c$ = 11.4414(5) | $V$ = 924.86(5) | ATL magnet ($J'/J$ = 0.40)<br>$T_N$ < 1.8 K | 23 |
| $Ba_3ReO_5Br_2$ | *Cmcm* (#63) | 5d$^1$ | $a$ = 5.87477(4)<br>$b$ = 14.1992(2)<br>$c$ = 11.60519(9) | $V$ = 968.07(4) | ATL magnet ($J'/J$ = 0.45)<br>$T_N$ < 1.8 K | 20 |
| $Ba_3WO_5Cl_2$ | *Cmcm* (#63) | 5d$^0$ | $a$ = 5.796(2)<br>$b$ = 13.825(2)<br>$c$ = 11.469(2) | $V$ = 919.01 | | 24 |
| $Pb_3ReO_5Cl_2$ | *Cmcm* (#63) | 5d$^1$ | $a$ = 5.59462(4)<br>$b$ = 13.26905(8)<br>$c$ = 10.97535(7) | $V$ = 814.33(3) | ATL magnet ($J'/J$ = 0.35)<br>$T_N$ < 1.8 K | 20 |
| $Pb_3WO_5Cl_2$ | *Cmcm* (#63) | 5d$^0$ | $a$ = 5.617(1)<br>$b$ = 13.067(3)<br>$c$ = 11.073(2) | $V$ = 812.81(3) | Mineral: pinalite | 14,15 |
| $Pb_3WO_5Br_2$ | *Cmcm* (#63) | 5d$^0$ | $a$ = 5.677(2)<br>$b$ = 13.323(3)<br>$c$ = 11.217(3) | $V$ = 848.4 | | 16 |
| $Pb_3MoO_5Cl_2$ | *Cmcm* (#63) | 4d$^0$ | $a$ = 5.59485(5)<br>$b$ = 13.1149(2)<br>$c$ = 11.0116(1) | $V$ = 807.98(1) | | 16 |

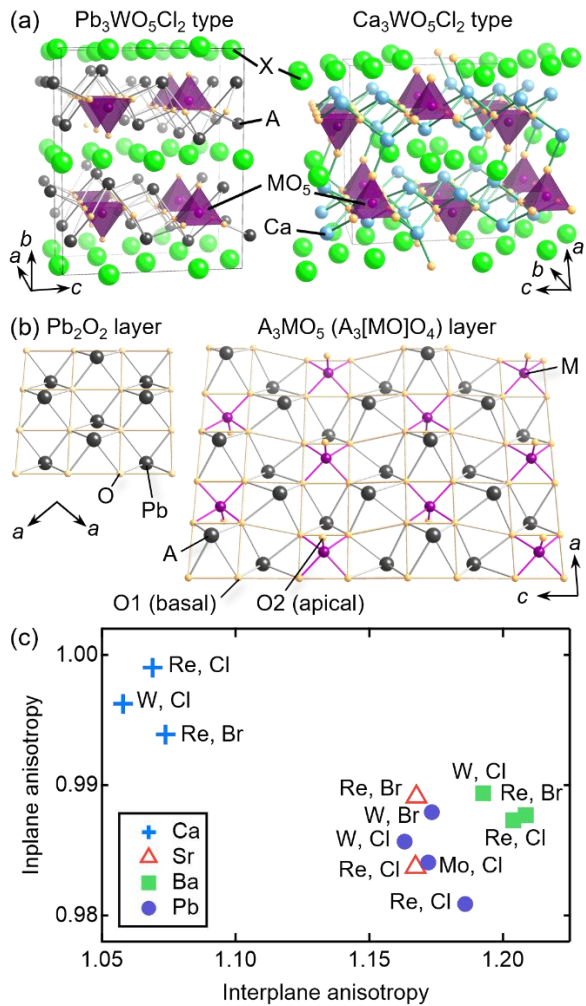

apical oxygen (O2) of the neighboring $A_3MO_5$ layer, resulting in a nine-coordinated monocapped square antiprism.

### 1.3 $Ca_3WO_5Cl_2$-TYPE STRUCTURE

When the A-site ion is $Ca^{2+}$, $A_3MO_5X_2$ compounds crystallize in the $Ca_3WO_5Cl_2$-type structure (Fig. 1a). The $Ca_3WO_5Cl_2$ and $Pb_3WO_5Cl_2$-type structures both have an $A_3MO_5$ layer, but they differ in the X layer. In the $Pb_3WO_5Cl_2$-type structure, the X layer forms a square lattice, but in the $Ca_3WO_5Cl_2$-type structure, the X layer is broken into one-dimensional (1D) slabs along the $b$ axis. This is because the Ca ions are too small to form a $Pb_3WO_5Cl_2$-type structure, and the coordination numbers of the $Ca^{2+}$ ions are reduced to five or six, compared to eight or nine for other $A^{2+}$ ions. Further, one of the three $Ca^{2+}$ ion in the $Ca_3WO_5Cl_2$-type structure is strongly bonded to the apical oxygen of the M-O unit in the adjacent $A_3MO_5$ layer, in addition to the four basal oxygens. This leaves no room for the X ions, and the X square lattice is, thus, split into slabs.

In the $Ca_3WO_5Cl_2$-type structure, the $M^{6+}$ ion also has a square pyramidal coordination, but the basal oxygen atoms form a trapezoid instead of a rectangle, as in the $Pb_3WO_5Cl_2$-type structure. This symmetry change affects the electronic properties, such as the optical and magnetic properties.

### 1.4 COMPARISON BETWEEN COMPOUNDS

Table 1 shows the lattice constants and unit cell volumes of $A_3MO_5X_2$ compounds. Changing the transition metal ion ($M^{6+}$) has little effect on the overall structure, and the volume change is less than 1%. This is reasonable because transition metal ions have similar ionic radii[13]. On the other hand, substituting the $X^-$ and $A^{2+}$ ions results in a significant structural change. To date, only materials with $Cl^-$ and $Br^-$ at the $X^-$ site have been synthesized. $Br^-$, which has a larger ionic radius than $Cl^-$, increases the unit cell volume ($V$) by more than 4%. The A-site cation has the largest effect on the unit cell volume: comparing $A_3ReO_5Cl_2$ compounds, when A = Pb, Sr, and Ba, $V$ is 10.6%, 12.4%, and 25.6% larger than that of A = Ca, respectively.

The inplane and interplane anisotropies of the $A_3MO_5$ layers, which are approximately $c/2a$ and $b/2a$, respectively, are compared in Fig. 1c (For A = Ca, the inplane and interplane anisotropies are $c/2b$ and $a/2b$). For α-PbO with a square lattice, the inplane anisotropy is one. All compounds are in the range $0.98 < c/2a < 1$, indicating that the square lattice formed by oxide atoms in the $A_3MO_5$ layer is close to a regular square lattice. The interplane distance is given by $a/2$ and $b/2$ for $Ca_3WO_5Cl_2$- and $Pb_3WO_5Cl_2$-type structures and varies from a minimum of 5.91 Å for $Ca_3WO_5Cl_2$ to 7.10 Å for $Ba_3ReO_5Br_2$. When normalized by the inplane lattice constants ($b$ and $a$), compounds with a $Ca_3WO_5Cl_2$-type structure have a smaller interplane anisotropy. This indicates that the $Pb_3WO_5Cl_2$-type structure is more two-dimensional (2D) than the $Ca_3WO_5Cl_2$-type structure.

For M = Re, $A_3MO_5X_2$ compounds with most combinations of A = Ca, Sr, Ba, and Pb and X = Cl and Br have been synthesized. Because Re, Mo, and W ions have similar ionic radii

**Figure 1.** (a) Crystal structures of $Pb_3WO_5Cl_2$-type in *Cmcm* (left) and $Ca_3WO_5Cl_2$-type in *Pnma* (right). Both structures commonly contain the $A_3MO_5$ layer. (b) Comparison between the [$Pb_2O_2$] layer in α-PbO (left) and the $A_3MO_5$ layer (right). Square lattice formed by oxide atoms are common in both layers, but the square lattice in the $A_3MO_5$ layer is corrugated. In the [$Pb_2O_2$] layer, Pb atoms are located above and below the center of the oxygen squares in a checkerboard pattern. In the $A_3MO_5$ layer, A atoms and M-O units are arranged in a similar manner to that of Pb atoms in the [$Pb_2O_2$] layer. (c) The inplane and interplane anisotropies of the $A_3MO_5$ layers in $A_3MO_5X_2$ compounds calculated from the lattice constants.

The $M^{6+}$ ion is surrounded by four basal oxygen atoms (O1) and one apical oxygen atom (O2), forming an $MO_5$ square pyramidal coordination. The M-O2 bond distance is shorter than the M-O1 bond distance, and the $M^{6+}$ ion is off the oxygen square plane. The unusual square pyramidal coordination of the $M^{6+}$ ion is likely influenced by the X ion on the opposite side of the apical oxygen.

The $A^{2+}$ ion has two sites: A1 and A2. A1 is coordinated by four basal oxide atoms (O1) and four X ions in a square antiprismatic geometry. A2 is also bonded to four basal oxide atoms and four X atoms, but in addition, it is bonded to the and electronegativities, compounds with M = Mo and W can probably be synthesized on further investigation. $Ba_3MoO_5Cl_2$ has been synthesized, although not in a pure phase[16]. Despite the limited number of ions that can have the 6+ valence state, it may be possible to synthesize materials with $Os^{6+}$, $Ru^{6+}$, and $Cr^{6+}$ at the M site in the future.

1.5 RELATED STRUCTURES

The $Pb_3WO_5Cl_2$-type structure is related to that of lead molybdenum oxychloride: $Pb_7MoO_9Cl_2$[16], known as the mineral parkinsonite[25]. This composition can be expressed as $Pb_3MoO_5[Pb_4O_4]Cl_2$, which can be regarded as a stacked structure of $A_3MO_5$, $Pb_2O_2$, and $X_2$ layers. Unlike the $Pb_3WO_5Cl_2$-type structure, in this structure, the Mo-O units randomly occupy the Pb sites[26]. It has been reported that analogous compounds having a composition of $Pb_7MO_9X_2$ (M = Mo, W; X = Cl, Br, I) [27] can be synthesized. By adjusting the synthetic conditions and selecting the constituent elements, it may be possible to obtain $A_7MO_9X_2$-type compounds with ideal stacking.

The $[Pb_2O_2]$ layer is structurally similar to the $[Bi_2O_2]$ layer in a series of materials with Aurivillius phases[28,29]. As in the Aurivillius compounds with alternating octahedral $nABO_3$ perovskite and $[Bi_2O_2]$ layers, the incorporation of perovskite octahedral layers between the $A_3MO_5$ layers is expected to further expand the structural diversity.

2 OPTICAL PROPERTIES

A remarkable physical property of CROC is its striking pleochroism (Fig. 2b). The color of crystals changes from green to reddish brown depending on the viewing direction (without polarizer), and the color changes from red to yellow to green depending on the polarization of the incident light (with polarizer). The heteroanionic coordination plays a crucial role in determining the optical properties.

2.1 VIOLATION OF LAPORTE RULE

*d-d* transitions, in which electrons are excited between d-orbitals upon photoabsorption, are responsible for the color of many transition metal compounds. *d-d* transitions also contribute to the pleochroism of CROC. In transition metal compounds, the energy levels of the five-fold degenerate d-orbitals of transition metal ions are split by the crystal field (so-called crystal field splitting (CFS)), as shown in Fig. 2a. The energy difference between the split d orbitals corresponds to the energy of visible light ranging from 1.5 to 3.5 eV. Therefore, by absorbing light with an energy matching the splitting, the electrons in the ground state d-orbital are excited, and the complementary color of the absorbed light becomes the color of the material. For example, in rubies, yellow–green light is absorbed by the *d-d* transition of $Cr^{3+}$ ions substituted for Al in $Al_2O_3$, resulting in a complementary color: red[30-32].

However, *d-d* transitions do not always occur. In particular, in the presence of an inversion center at the transition metal site, the *d-d* transition is forbidden, and this is known as the Laporte selection rule[33]. For monoanionic compounds, the coordination is mostly tetrahedral or octahedral depending on the relative ionic sizes of the transition metal ions and anions. Many oxides have centrosymmetric octahedral coordination, in which there is an inversion center at the transition metal site. Thus, the *d-d* transition is forbidden, and no light absorption occurs; thus, there is no color.

To violate the Laporte rule and achieve intense optical absorption, coordination without an inversion center at the transition metal site is necessary. A good example is the recently discovered vibrant blue pigment YInMn Blue, which consists of $YIn_{1-x}Mn_xO_3$[34]. In this material, the transition metal ion responsible for optical absorption, $Mn^{3+}$, has a trigonal bipyramidal coordination without an inversion center at the transition metal site, which allows *d-d* transitions. This unique coordination enables intense absorption of visible light, resulting in vibrant pigments. Other inorganic pigments with various colors have also been produced, mainly focusing on trigonal bipyramidal coordination[35].

Crucially, heteroanionic compounds have the advantage of violating the Laporte rule and achieving intense optical absorption. In heteroanionic coordination, multiple species of anions coordinate to the transition metal ion, making it easier to achieve low-symmetry noncentrosymmetric coordination. In particular, replacing one vertex of the $MO_6$ octahedron with one chloride ion, as shown in Fig. 2a, results in the loss of the inversion center at the transition metal site and, thus, allows *d-d* transitions. In CROC, the $Re^{6+}$ ion is surrounded by a square pyramidal coordination of five oxide atoms and one chlorine atom (Fig. 2a), which allows light absorption via the *d-d* transition.

2.2 COMPLEX CFS

Heteroanionic coordination plays an important role not only in the intensity of light absorption but also in the energy of light absorption. In the case of octahedral coordination of oxides, CFS splits the d-orbital into a doubly degenerate $e_g$ and a triply degenerate $t_{2g}$ symmetry orbitals. Therefore, the only energy at which the transitions occur upon photoabsorption lies between $e_g$ and $t_{2g}$ levels (Fig. 2a).

In contrast, in CROC, the $e_g$ and $t_{2g}$ orbitals are further split by heteroanionic coordination: the divalent apical oxide atom in the *z* direction is replaced by a monovalent chloride ion with a longer interatomic distance, so that the $d_{z^2}$ and $d_{x^2-y^2}$ orbitals in the $e_g$ symmetry orbitals are significantly separated and the $d_{yz}/d_{zx}$ and $d_{xy}$ orbitals of $t_{2g}$ symmetry are also split. Furthermore, the degenerate $d_{yz}/d_{zx}$ orbitals split slightly into two linear combinations, $d_{xz-yz}$ and $d_{xz+yz}$, as a result of the deformation of the basal square of oxide atoms to a trapezoid because of the chemical pressure from $Ca^{2+}$ ions. The energies from the ground state $d_{xy}$ orbital to the second highest $d_{z^2}$ orbital and the highest $d_{x^2-y^2}$ orbital correspond to the energies of red and green light, respectively. These complex energy levels lead to the absorption of light at various wavelengths, resulting in the pleochroism of different colors depending on the viewing direction. This complex CFS has been observed in other heteroanionic compounds and are often discussed in relation to the spin states[36].

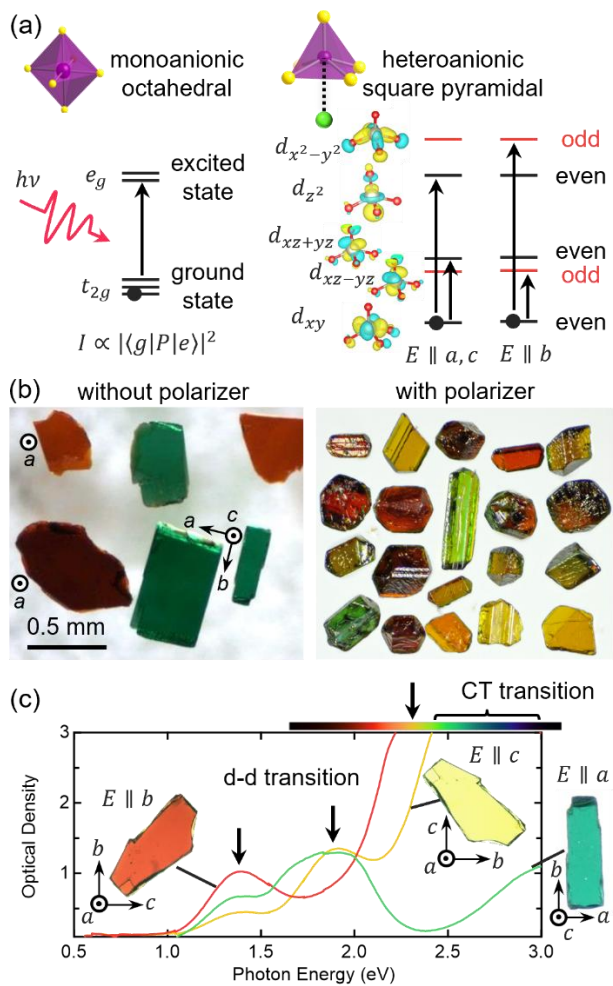

**Figure 2.** (a) Comparison of crystal field splitting of d-orbitals between monoanionic octahedral coordination (left) and heteroanionic square-pyramidal coordination of CROC (right). The parity of orbitals with respect to the mirror plane perpendicular to the $b$ axis is shown on the right. (b) Photo of single crystals of CROC taken without (left) and with (right) polarizer. (c) Optical density spectra for CROC with linearly polarized incident light parallel to each axis. The corresponding photos of crystals are shown. The arrows indicate the absorption peaks corresponding to $d$-$d$ transitions. The intense absorption for $E \parallel b$ and $c$ above 2.4 eV is attributed to the charge transfer transition.

## 2.3 MECHANISM OF PLEOCHROISM

As shown in Fig. 2c, CROC is trichroic, meaning that when the polarization of the incident light is parallel to the $b$, $c$, and $a$ axes, the crystal colors are red, yellow, and green, respectively. The optical density spectra for the linearly polarized light of CROC show absorption peaks originating from $d$-$d$ transitions and broad and intense absorption originating from charge transfer (CT) transitions. As mentioned above, $d$-$d$ transitions are allowed in CROC because of the absence of local inversion symmetry at the transition metal site. CT transitions, which are generally Laporte-allowed transitions between ligands and transition metal ions, exhibit intense optical absorption at higher energies than the $d$-$d$ transitions. In addition, $d$-$d$ transitions tend to have narrower absorption peaks because the bands of the d-orbitals are sharper than those of the p-orbitals of the ligand. Based on these characteristics, the absorption peaks at 1.4, 1.9, and 2.3 eV can be assigned as $d$-$d$ transitions, and continuous intense absorption above 2.4 eV for $E \parallel b$ and $c$ and above 3.0 eV for $E \parallel a$ can be assigned as CT transitions. The pleochroism of CROC is attributed to the strong polarization dependence of the three d–d and CT transitions.

First, the polarization dependence of the $d$-$d$ transition can be explained by considering the d-orbital arrangement split by the CFSs and the optical selection rule for electronic transitions. According to Fermi's golden rule, the probability of an electronic transition, i.e., the intensity of optical absorption, is proportional to the transition dipole moment, as follows:

$$I \propto |\langle g|P|e\rangle|^2$$

where $e$, $g$, and $P$ denote the excited state, ground state, and the dipole moment of light, respectively. Electronic transitions are allowed when the transition dipole moment is finite. This is when the combination of the parities of the three components is even; in the CROC case, we consider the parities of the three components with respect to a mirror plane perpendicular to the $b$ axis, which is the local symmetry of the Re site. The parity of each d orbital is shown in Fig. 2a, where the ground state $d_{xy}$ orbital ($g$) is even, the excited states ($e$) $d_{xz+yz}$ and $d_{z^2}$ are even, and $d_{xz-yz}$ and $d_{x^2-y^2}$ are odd. The parity of the light ($P$) is different depending on the polarization; light polarized along the $b$ axis ($E \parallel b$) has odd parity with respect to the mirror plane, and light polarized along the $a$ or $c$ axis ($E \parallel a$ or $c$) has even parity.

Because the ground state ($g$) is a $d_{xy}$ orbital with even parity, electronic transitions are allowed only when $e$ and $P$ have the same parity. In other words, when $P$ is odd ($E \parallel b$) transitions to the $d_{xz-yz}$ and $d_{x^2-y^2}$ orbitals where $e$ is odd occur (Fig. 2a). On the other hand, when $P$ is even ($E \parallel a$ or $c$), transitions to the $d_{xz+yz}$ and $d_{z^2}$ orbitals with even parity occur. The $d$-$d$ transitions at 1.4 and 2.3 eV observed for $E \parallel b$ correspond to excitations to the $d_{xz-yz}$ and $d_{x^2-y^2}$ orbitals, while the transitions at 1.4 and 1.9 eV for $E \parallel a$ and $c$ correspond to excitations to the $d_{xz+yz}$ and $d_{z^2}$ orbitals. The fact that the $d$-$d$ transitions to the $d_{xz-yz}$ and $d_{xz+yz}$ orbitals are observed at almost the same energy implies that the energy splitting between the two orbitals is small, as expected from the small trapezoidal distortion at the base of the ReO$_5$ square pyramid.

The difference in color between light polarized along the $a$ and $c$ axes comes from the energy difference in the CT transitions: in CROC, the ReO$_5$ square pyramids are aligned along the $a$ axis, and light polarized along the $a$ axis induces transitions from the $p_z$ orbital of the apical oxygen to the Re $d_{z^2}$ orbital. On the other hand, light polarized along the $c$ axis induces transitions from p orbitals of the basal oxygen and the other Re 5d orbitals ($d_{xy}$, $d_{xz-yz}$, $d_{xz+yz}$, and $d_{x^2-y^2}$). It is possible that the strong hybridization of $p_z$ and $d_{z^2}$ is the origin of the higher CT transition energies for $E \parallel a$ (above 3.0 eV) than those for $E \parallel c$ (above 2.4 eV). As described above, heteroanionic coordination, which violates Larporte rule and induces unique CFS, plays an important role in the pleochroism exhibited by CROC.

2.4 OUTLOOK AND OTHER OPTICAL PROPERTIES

The pleochroism of CROC is very sensitive to the coordination of the $Re^{6+}$ ions. Therefore, the color can be controlled by replacing the ions at the A and X sites. In fact, our preliminary experiments have shown that replacing the A site with Sr and Ba increases the distance between $Re^{6+}$ and the ligands and weakens the CFS, resulting in redshifted spectra. By growing single crystals of other $A_3MO_5X_2$ compounds, materials that exhibit pleochroism with different color variations from CROC can be synthesized. Therefore, violating the Laporte rule using heteroanionic coordination and enabling intense light absorption may be an effective strategy for developing new pigments. New pigments could be obtained by using a square pyramid and other heteroanionic coordination compounds.

Even in heteroanionic compounds without d-electrons, optical properties that are not observed in monoanionic compounds can be expected. For example, $Ca_3WO_5Cl_2$ with $W^{6+}$ ions in the $5d^0$ electronic configuration has been investigated for its luminescence properties; notably, $Ca_3WO_5Cl_2$ exhibits blue luminescence under UV light illumination because of the CT transition between $W^{6+}$ and the ligand ions[22]. Compared with the monoanionic compound $Ca_3WO_6$, the luminescence band is redshifted by 0.36 eV. In contrast to monoanionic compounds, which have limited variation in the coordination polyhedra, heteroanionic compounds allow fine control of the CFS, which may also enable control of the luminescence properties. CT fluorescence of several heteroanionic tungsten compounds has already been reported[37-39]: $BaWO_2F_4$, which forms a distorted octahedron of cis-$[WO_2F_4]^{2-}$, has shown green luminescence with high quantum yield[39]. I hope that the luminescence properties of $A_3MO_5X_2$ compounds other than $Ca_3WO_5Cl_2$ will be investigated to discover desirable and highly efficient phosphors and scintillators.

In $A_3MO_5X_2$ compounds, the crystal structure has global inversion symmetry. However, heteroanionic coordination without a local inversion center sometimes leads to noncentrosymmetric crystal structures. Such structures cause nonlinear optical properties such as second-harmonic generation. Heteroanionic compounds have attracted increasing attention as new nonlinear optical materials[40,41]. In the future, nonlinear optical properties will also be an important direction to explore in $A_3MO_5X_2$ compounds.

3 MAGNETIC PROPERTIES

The 5d electrons of the Re ions exhibit unique quantum magnetism. CFS arising from heteroanionic coordination gives rise to a spin state distinct from that of other 5d magnets and a low dimensionality derived from selective orbital occupation.

3.1 STABILIZATION OF THE SPIN-1/2 STATE

In many 3d transition-metal-based magnets, the orbital angular momentum is quenched; therefore, the spin quantum number $S$ is a good quantum number without considering the spin–orbit interactions (SOIs). In contrast, 5d electrons usually have strong SOIs, forming an electronic state in which the spin and orbital angular momenta are entangled[42-44].

The electronic state of the 5d orbital is described by the spin–orbit-entangled $J_{eff} = 3/2$ states in compounds with octahedrally coordinated $Re^{6+}$ ions in a $5d^1$ electronic configuration (Fig 3a). The effective orbital angular momentum $l_{eff} = 1$ derived from the triply degenerate $t_{2g}$ orbitals coupled with the spin angular momentum $s = 1/2$ renders the total angular momentum $J_{eff} = 3/2$ a good quantum number. The cancellation between the spin momentum and effective orbital momentum results in the total magnetic moment $M$ being significantly reduced from the value expected for $S = 1/2$[45]. The effective magnetic moment expected for $S = 1/2$ is 1.73 $\mu_B$, but only 0.68–0.8 $\mu_B$ has been reported for $A_2BReO_6$ (A = Ba, Sr; B = Mg, Ca, Cd) with $Re^{6+}$ ions[46-49].

The electronic state of the $Re^{6+}$ ion in CROC is distinct from that of monoanionic 5d magnets. Heteroanionic coordination is advantageous for stabilizing the spin-1/2 state instead of the spin–orbit-entangled $J$ state expected for 5d electrons. As discussed for the optical properties, in CROC, the degeneracy of the 5d orbitals of the Re ion is completely lifted by the CFSs of heteroanionic coordination compounds. First-principles calculations show a 1.4 eV energy separation between the ground state $d_{xy}$ orbital and the first excited state[18] (Fig. 3a). Thus, the orbital angular momentum is quenched. Although the SOI is as large as 0.5 eV in 5d electron systems, the effect of SOI is negligible owing to the large CFSs. The effective magnetic moments obtained from the Curie–Weiss fit to the magnetic susceptibility of CROC at high temperatures range from 1.56 to 1.65 $\mu_B$, close to the values expected for spin-1/2 (1.73 $\mu_B$)[50]. In addition, the $g$-values obtained from electron spin resonance (ESR) experiments are almost isotropic (1.85–1.92) and close to 2 for the spin-only value. Based on these results, the exchange interaction was isotropic, and CROC was studied as a model of an ideal Heisenberg spin quantum magnet with spin-1/2.

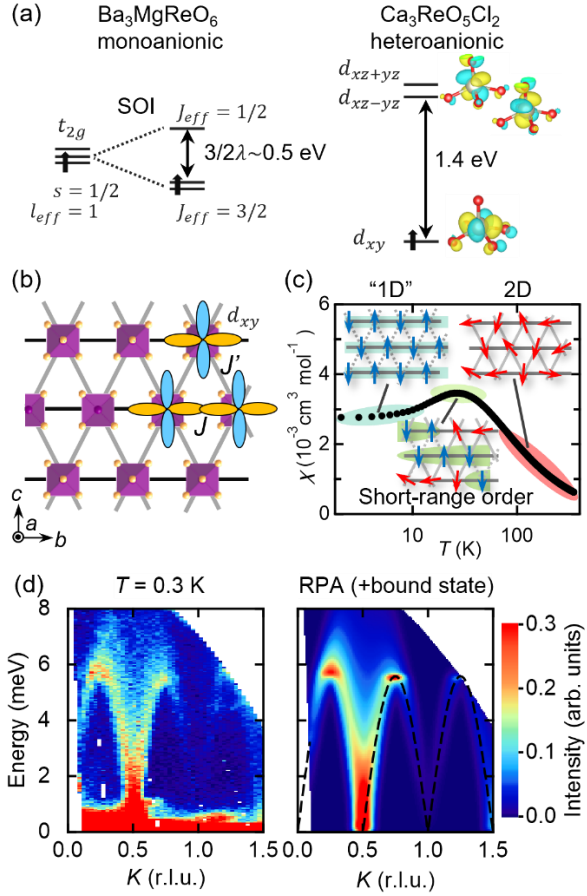

**Figure 3.** (a) Comparison of electronic states of magnetic ion with 5d[1] electronic configuration between monoanionic octahedral and heteroanionic square pyramidal coordination. (b) Arrangement of $d_{xy}$ orbitals in CROC forming anisotropic triangular lattice on the $bc$-plane. (c) Schematic of dimensional reduction in an ATL antiferromagnet shown with the magnetic susceptibility of CROC. The spin alignments of corresponding temperature range are also shown. (d) Inelastic neutron scattering spectra of CROC measured at 0.3 K (left) and simulated by applying random-phase approximation (RPA) to antiferromagnetically coupled spin-1/2 chains (right).

## 3.2 ANISOTROPIC INTERACTION ORIGINATING FROM ORBITAL STATE

Orbital splitting due to heteroanionic crystal fields affects not only the spin states but also the magnetic interactions between the spins. Because CFSs cause electrons to occupy specific orbitals, the magnetic interactions are also anisotropic, reflecting orbital anisotropy.

In CROC, the Re coordination polyhedron does not share ligands with other polyhedra; the distances between the Re ions are 5.5661(1) and 6.3989(3) Å within the $bc$ planes and 5.5515(3) Å with Re ions in different $bc$ planes[19]. Considering the Re-Re distances, one would expect the interaction between the $bc$ planes ($J''$) to be the strongest. However, first-principles calculations estimate that $J''$ is three orders of magnitude smaller than the strongest interaction $J$ ($J:J''$ = 1:0.0007) [19]. These highly anisotropic interactions are attributed to orbital occupancy, in which electrons occupy the lowest-energy $d_{xy}$ orbitals lying on the $bc$ plane. Therefore, the orbital overlap in the $J''$ direction was significantly small, and the magnetic interaction was negligible despite the short Re-Re distance. Therefore, CROC can be effectively regarded as a 2D magnet.

In the $bc$ plane, the Re[6+] ions form a triangular lattice, which is expected to exhibit geometrical frustration. This triangular lattice is not a regular triangular lattice but a triangular lattice stretched in the $c$-axis direction. In addition, alignment of the $d_{xy}$ orbitals causes highly anisotropic magnetic interactions. Along the $b$-axis, one pair of electron lobes of the $d_{xy}$ orbital (orange in Fig. 3b) overlaps with those of two neighboring $d_{xy}$ orbitals, which leads to the dominant intrachain antiferromagnetic exchange, i.e., $J$. On the other hand, another pair of orbital lobes (blue in Fig. 3b) is arranged in a staggered manner between adjacent chains with a smaller overlap, resulting in the weak interchain exchange of $J'$. According to first-principles calculations and subsequent high-field ESR experiments, the $J'/J$ anisotropies were estimated to be 0.295[19] and 0.254[51], respectively. It should be noted that direct exchange interactions are sufficiently large even though the transition metals are separated from each other by more than 5 Å.

## 3.3 MODEL SYSTEM FOR ATL MAGNET

CROC and related materials have recently attracted great attention as model materials for frustrated magnetism on anisotropic triangular lattices (ATL) because of the ideal spin-1/2 state and the two-dimensionality derived from the orbital arrangement[19,23].

Theoretically, the spin-1/2 Heisenberg antiferromagnet on an ATL gives rise to intriguing frustration-induced phenomena, called "*dimensional reduction*" [52,53]. The ATL is composed of an intrachain interaction $J$ and an interchain interaction $J'$ (Fig. 3b). ATL with $J'/J$ = 1 corresponds to a regular triangular lattice, whereas in the limit of $J'/J$ = 0 it is identical to a decoupled 1D spin chain. Even with a moderately large interchain interaction ($0 < J'/J \lesssim 0.6$), the competing interactions $J'$ between the chains markedly reduce interchain correlations, and the 2D magnet behaves as an isolated 1D spin chain[54]. As a result, the Tomonaga–Luttinger liquid (TLL) state, which is a characteristic of 1D magnets, is realized at low temperatures.

Many studies have been conducted on $Cs_2CuCl_4$, which has a $J'/J$ anisotropy of 0.3. In $Cs_2CuCl_4$, a TLL-like state was observed at low temperatures as a continuous excitation of fractionalized spinons carrying spin-1/2 in inelastic neutron scattering experiments[55,56]. Although the TLL state is expected to be in the ground state, a spiral magnetic order is formed in $Cs_2CuCl_4$ at a Néel temperature ($T_N$) = 0.62 K as a result of Dzyaloshinskii–Moriya interactions (DMIs) and interplane interactions. The frustration parameter, $J/T_N$, which indicates the strength of the suppression of magnetic ordering by frustration, is approximately 7, suggesting a strong influence of interactions other than $J$ and $J'$. CROC with similar anisotropy also shows a spiral magnetic order at $T_N$ = 1.13 K. However, CROC with high two-dimensionality and ideal spin-1/2 state shows stronger suppression of magnetic order than $Cs_2CuCl_4$, with $J/T_N$ = 35. In addition, CROC has a large magnetic interaction $J$, which allows the properties of the ATL to be investigated over a wide temperature range. Therefore, CROC has attracted considerable attention as a better model material for ATL magnets.

The expected dimensional reduction in CROCs has been revealed by the magnetic susceptibility and heat capacity[19]: from the Curie–Weiss fit, the Weiss constant was obtained to be 40 K, which indicates that antiferromagnetic interactions are strong enough to induce magnetic ordering below 40 K in the absence of frustration, but no clear magnetic ordering occurred until $T_N$ = 1.13 K. In the magnetic susceptibility measurements, a broad peak indicating short-range order was observed at approximately 20 K (Fig. 3c). This temperature dependence of the magnetic susceptibility is not fitted by a 2D model but can be reproduced well by a 1D model. Furthermore, a large Sommerfeld coefficient was observed for the heat capacity at low temperatures, even though CROC is an insulator. The observed large Sommerfeld coefficient corresponding to spinon excitation is a characteristic of 1D magnets. Subsequently, continuous spinon excitation, which is clear evidence of the TLL state, was observed at low temperatures in inelastic neutron scattering[50] (Fig. 3d) and Raman spectroscopy experiments[57]. Based on these results, the low-temperature state of CROC can be regarded as a 1D magnet.

The interchain interaction $J'$ gives rise to a feature in ATL magnets that distinguishes them from 1D spin chains; because of $J'$, spinon hopping between neighboring chains is prohibited. Instead, $J'$ causes the formation of a bound spinon pairs (triplons)[52,53]. These triplons propagate coherently between the chains. In CROC, the interchain hopping of triplons was captured by inelastic neutron scattering experiments as sharply dispersive modes perpendicular to the chain[50]. Magnetic excitation of both spinons and triplons was also observed in the Raman scattering data[57]. The observed temperature dependence of the spinon excitations is different from that expected for 1D spinons and may be a characteristic phenomenon of dimensional reduction. Thus, novel phenomena originating from dimensional reduction, which have not been experimentally verified owing to the lack of model materials, are now being revealed.

3.4 OUTLOOK

Research is currently focusing on CROC, for which single crystals several millimeters in size can be obtained. However, analogous compounds with a $Pb_3WO_5Cl_2$-type structure should be ideal ATL magnets because of the weaker interplane interactions and prohibited DMIs owing to crystal symmetry. Thus, a spin-liquid state that does not exhibit long-range magnetic ordering, even at the lowest temperatures, may be realized.

Recent high-field ESR measurements revealed that the spiral magnetic order in CROC is stabilized by uniform DMIs[51]. Because the incommensurate magnetic order stabilized by DMIs is expected to host interesting phenomena such as magnetoelectric effects and magnetic skyrmions, the spiral magnetic order in CROC has garnered interest.

As discussed above, the substitution of A and X ions gives rise to anisotropic deformation of the crystal structure (Fig. 1c). Therefore, the magnetic anisotropy $J'/J$ also varies from 0.25 in CROC to 0.45 in $Ba_3ReO_5Br_2$[20] (Table 1). The change in the magnetic ground state depending on magnetic anisotropy is also of interest.

Orbital degeneracy tends to be reduced in heteroanionic compounds by CFSs with low-symmetry heteroanionic coordination. Thus, the orbital angular momentum is quenched and ideal spin-1/2 quantum magnetism is easily achieved. This is suitable for studying quantum magnetism.

Anisotropic magnetic interactions, which reflect orbital anisotropy, also occurs in many heteroanionic compounds because electrons occupy a specific d orbital. The anisotropy of these interactions may be a source of a new low-dimensional magnetism that is difficult to realize in monoanionic compounds. Therefore, heteroanionic compounds represent a good platform for the development of new low-dimensional magnets.

Heteroanionic coordination often leads to low dimensionality in the crystal structure, and $A_3MO_5X_2$ compounds have layered structures, with oxygen and chlorine forming different layers. A tendency to adopt layered structures was observed in heteroanionic compounds containing oxygen and heavy anions such as Cl, Br, S, and Se. For example, in $Sr_2CuO_2Cl_2$, the parent compound of high-$T_c$ cuprate superconductors, four oxide and two Cl atoms are coordinated to Cu, and Cl selectively occupies the apical oxygen position, forming a layered structure with high two-dimensionality. Owing to its high two-dimensionality, $Sr_2CuO_2Cl_2$ has been studied as a good model material for 2D square lattice quantum magnet[58,59]. Although heteroanionic compounds have already been recognized as a good platform for quantum magnetism research, and many studies have been devoted to them, future material searches will lead to the discovery of model magnets and associated quantum magnetic phenomena.

4. RELATED MIXED-ANION COMPOUNDS

Several compounds are known in which one halogen and five oxygen atoms coordinate to a metal atom in a square pyramidal configuration, similar to CROC. The most diverse compounds are $A_2M^{3+}O_3X$, which are analogs of $K_2NiF_4$-type layered perovskites, where A = Sr and Ca, M = Sc, V, Mn, Fe, Co, and Ni, and X = F, Cl, and Br [36,60–67]. See review article for oxyfluorides with related layered perovskite structures[68]. The magnetic properties of the materials were also investigated. In $Sr_2MO_3Cl$ (M = Mn, Co, Ni), a broad peak in the magnetic susceptibility characteristic of low-dimensional magnets has been observed[36,62,67], suggesting that the crystal structure and orbital arrangement are responsible for low-dimensional magnetism. In $Sr_2CoO_3F$, spin crossover from high-spin to low-spin states has been reported to occur upon the application of pressure[69], which is considered a unique phase transition arising from the change in heteroanionic coordination.

Although the magnetism of $A_2MO_3X$ has been extensively investigated, the optical properties have been neglected. However, it has been reported that $Sr_2FeO_3Cl$ and $Sr_2MnO_3Cl$ exhibit various colors: reddish brown[63] and dark green[70], respectively, possibly because of the intense optical absorption in the visible light region as a result of the square pyramid coordination, which allows *d-d* transitions. The optical properties of these materials will be investigated in future studies.

There are already several review articles on other heteroanionic compounds such as oxyhalides[71,72], oxychalcogenides, oxypnictides[73], oxyhydrides[74], and chalcohalides[75]. Similar to the Pinalite family, many heteroanionic compounds, such as oxyhalides[72] and oxypnictides[73], take a layered structure, while unique crystal chemistry has been revealed in compounds such as chalcohalides[75] and oxyfluorides[71]. Heteroanionic compounds have also been reviewed in terms of their applications: nonlinear optics[76], semiconductors, photocatalysts, batteries, and superconductors[72,75,77]. Recent attempts to capture design strategies lead to the on-demand synthesis of designer heteroanionic materials[78]. By incorporating new synthetic ideas, it may be possible to expand the Pinalite family.

## 5. CONCLUSIONS

In conclusion, the recent progress and future prospects of research on the crystal structures, optical properties, and magnetism of CROC, which was discovered in 2017, and related $A_3MO_5X_2$ compounds have been described. Heteroanionic coordination in heteroanionic compounds strongly affect the d-orbitals of transition metal ions, resulting in unique electronic properties. Since the electronic features of the Pinalite family described in this review (active d-d transitions, complex orbital splitting, and magnetic anisotropy) are universally inherent in many heteroanionic compounds, I hope that this work can stimulate more investigations on optical and magnetic properties of heteroanionic materials. Research on heteroanionic compounds has been active in recent years, and new physical properties and functions have been discovered along the way. Further development is expected in the future.

## AUTHOR INFORMATION

**Corresponding Authors**

*dhirai@nuap.nagoya-u.ac.jp

**Funding Sources**

This work was partly supported by Japan Society for the Promotion of Science (JSPS) KAKENHI Grants of Numbers JP20H01858, JP22H04462 (Quantum Liquid Crystals), and JP23H04860 and the Toyota Riken Scholar Program from Toyota Physical and Chemical Research Institute.

**Notes**

The authors declare no competing financial interest.

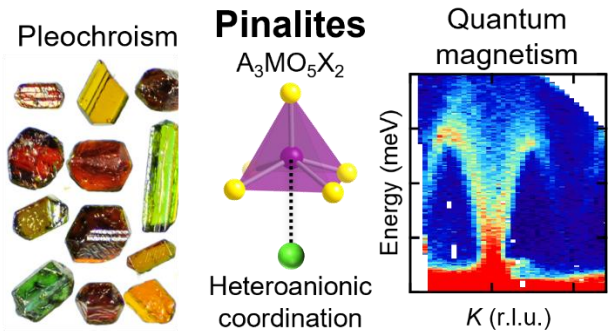

Synopsis

The structure, optical properties, and quantum magnetism of pinalite, a heteroanionic compound with a chemical formula of $A_3MO_5X_2$, are reviewed. The heteroanionic coordination of transition metal ion is the key to the pleochroism and quantum magnetism.

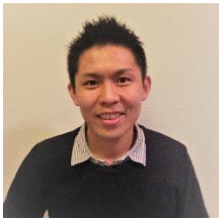

Biography

Daigorou Hirai is an Associate Professor in the Department of Applied Physics at Nagoya University. He received his PhD in materials science from the University of Tokyo (2011). Afterwards, he worked as a postdoctoral fellow in the Department of Chemistry at Princeton University from 2011 to 2013 and in the Department of Physics at the University of Tokyo from 2013 to 2015. Thereafter, he served as a Research Associate at the Institute for Solid State Physics, the University of Tokyo from 2015 to 2022. His research interests include the synthesis, and the structural chemistry of transition-metal based oxides and heteroanionic compounds, in relation to their electronic (magnetic and optical, and superconducting) properties.